\PassOptionsToPackage{table}{xcolor}
\documentclass[unnumsec,webpdf,contemporary,large]{oup-authoring-template}%

\makeatletter

\def\ps@opening{%
  \let\@oddhead\@empty
  \let\@evenhead\@empty
  \let\@oddfoot\@empty
  \let\@evenfoot\@empty
}

\AtBeginDocument{%
  \gdef\@evenhead{%
    \hbox to \textwidth{%
      \sffamily\fontsize{8bp}{10bp}\selectfont
      {\fontseries{m}\fontsize{9bp}{10bp}\selectfont\thepage}%
      \hfill}}%
  \gdef\@oddhead{%
    \hbox to \textwidth{%
      \hfill
      \sffamily\fontsize{8bp}{10bp}\selectfont
      {\fontseries{m}\fontsize{9bp}{10bp}\selectfont\thepage}}}%
}

\makeatother

\graphicspath{{Fig/}}

\theoremstyle{thmstyleone}%

\theoremstyle{thmstyletwo}%
\theoremstyle{thmstylethree}%

\usepackage{amsmath,amssymb,mathtools,bm,bbm}
\usepackage{tabularx}
\usepackage[table]{xcolor}
\usepackage{array}
\usepackage{booktabs}

\definecolor{seratiogray}{HTML}{F3F3F3}
\newcolumntype{G}{>{\columncolor{seratiogray}}c}

\newcolumntype{Y}{>{\raggedright\arraybackslash}X}

\newcommand{\Var}{\operatorname{Var}}

\newcommand{\trans}{\mathsf{T}}

\newcommand{\bbeta}{\bm{\beta}}
\newcommand{\btheta}{\bm{\theta}}

\begin{document}

\journaltitle{Journal of the Royal Statistical Society B}
\DOI{DOI added during production}
\copyrightyear{YEAR}
\pubyear{YEAR}
\vol{XX}
\issue{x}
\access{Published: Date added during production}
\appnotes{Paper}

\firstpage{1}


\title[Survey-robust Variance Estimation in GAMLSS]{Survey-robust uncertainty quantification in generalised additive models for location, scale, and shape}

\author[1,$\ast$]{Dennis Meurer}
\author[1]{Timo Adam}

\address[1]{\orgdiv{Department of Business Administration and Economics}, \orgname{Bielefeld University}, \orgaddress{\street{Universitätsstraße 25}, \postcode{33615}, \state{North Rhine-Westphalia}, \country{Germany}}}

\corresp[$\ast$]{Corresponding author. \href{email:dennis.meurer@uni-bielefeld.de}{dennis.meurer@uni-bielefeld.de}}

\abstract{Conventional mean-only regression models are often too restrictive for the analysis of complex survey data, where interest frequently extends beyond the conditional mean to other aspects of the response distribution. Generalised additive models for location, scale and shape (GAMLSS) provide a flexible framework by allowing all distributional parameters to depend on covariates. However, existing model-based and model-robust standard errors fail to account for complex survey designs and can substantially underestimate the variability of regression parameter estimates. We propose a linearisation-based sandwich variance estimator that incorporates the survey design while remaining computationally efficient. Using a simulation study based on synthetic survey data, we demonstrate that the proposed estimator provides accurate standard error estimates and reliable confidence interval coverage for all distributional parameters, while offering a computationally efficient alternative to replication-based methods. We further illustrate the practical utility of the approach through a re-analysis of data from the 2019/20 Rwanda Demographic and Health Survey (DHS).}


\keywords[Abbreviations]{GAMLSS, Generalised Additive Models for Location, Scale and Shape; DHS, Demographic and Health Survey; PSU, primary sampling unit; PPS, probability proportional to size; SE, standard error; ICC, intraclass correlation coefficient; FAMD, factor analysis for mixed data}

\boxedtext{Key Messages}{
\begin{itemize}
\item Conventional GAMLSS standard errors can substantially underestimate uncertainty when models are fitted to survey data with clustering and unequal sampling weights.
\item We develop a linearisation-based, survey-robust sandwich estimator that combines stacked GAMLSS score contributions with first-stage survey design information.
\item In simulation studies, the proposed estimator closely matched empirical repeated-sampling variability and achieved near-nominal centred Wald coverage across a range of challenging complex-survey settings.
\item Applied to data from the 2019/20 Demographic and Health Survey (DHS) in Rwanda, the proposed estimator calls into question several previously reported associations between childhood stunting and socio-economic factors that were supported by conventional model-based standard errors.
\end{itemize}}

\maketitle

\section{Introduction}

As survey data become increasingly complex and widely available, mean-only regression models are often too restrictive. In many applications, researchers are interested not only in the conditional mean but also in other features of the response distribution, such as dispersion, skewness, and tail behaviour. Generalised additive models for location, scale, and shape (GAMLSS; \citealp{rigbyGeneralizedAdditiveModels2005}) provide a flexible framework for such distributional regression by allowing multiple parameters of the response distribution to depend on covariates. This capability is particularly valuable for the analysis of complex survey data, where understanding heterogeneous risks and subgroup differences across the entire response distribution is often of primary interest.

Despite the potential of distributional regression for survey analysis, its integration with complex survey inference remains limited. The small body of applied work using GAMLSS with complex survey data has relied primarily on replication-based variance estimation; for example, \citet{yamadaDifferencesMagnitudeRate2019} employ bootstrap methods. More fundamentally, existing GAMLSS software does not natively accommodate key features of complex survey designs, including sampling weights, stratification, and clustering. As a result, authors have explicitly acknowledged this as an important limitation. For example, \citet{xiEstablishingInternationalBlood2015} note that GAMLSS do not account for the sampling weights required for analysing complex survey data and conclude that \textit{``this issue should be acknowledged as an unsolved limitation''}. Similarly, \citet{hernandez-vasquezBloodPressureReference2026} report that omitting survey weights and design features due to limitations of the \texttt{gamlss} package may lead to underestimated standard errors.

A general design-based variance framework for distributional regression is therefore currently lacking. Although the \texttt{gamlss} package in \texttt{R} provides both naive model-based standard errors and model-robust sandwich estimators following \citet{huberBehaviorMaximumLikelihood1967a} and \citet{whiteHeteroskedasticityConsistentCovarianceMatrix1980}, neither accounts for the complex sampling designs commonly used in survey data. As we demonstrate in this paper, both approaches can substantially underestimate standard errors, resulting in anticonservative statistical inference. To address this limitation, we propose a survey-robust, linearisation-based sandwich variance estimator for GAMLSS. Building on the framework of \citet{binderVariancesAsymptoticallyNormal1983}, our approach extends design-based variance estimation to distributional regression by exploiting stacked score equations, thereby accounting for the joint estimation of multiple distributional parameters and their covariance. The resulting estimator yields design-consistent standard errors under complex survey designs, while remaining applicable to the broad range of response distributions supported by GAMLSS. Unlike replication-based methods, it requires only a single model fit, making it a computationally efficient alternative.

The remainder of the paper is organised as follows. We begin by reviewing GAMLSS and its unit score contributions before embedding the model in the implicit-parameter framework of \citet{binderVariancesAsymptoticallyNormal1983}. We then derive the corresponding bread matrix, apply a Taylor linearisation to the estimator, and combine it with a one-stage clustered survey-design meat matrix to obtain the proposed variance estimator. Next, through a series of ten increasingly complex simulation studies, we show that the proposed estimator provides reliable inference under complex sampling, whereas both naive model-based and model-robust estimators substantially underestimate standard errors. We further demonstrate that, while replication-based methods achieve comparable statistical performance, the proposed estimator is considerably more computationally efficient. Finally, we revisit the analysis of \citet{meurerDeterminantsChildhoodMalnutrition2024} on the determinants of childhood stunting in Rwanda and show that, although the main substantive conclusions remain unchanged, several associations are no longer statistically significant when inference properly accounts for the survey design. An \texttt{R} implementation of the proposed method is available online\footnote{See \url{https://github.com/dmeurer94/survey-robust-gamlss}.}.

\section{Methods}\label{sec4}

In this section, we present the proposed methodology. We first introduce the GAMLSS framework using standard notation and formulate it within the finite-population framework of \citet{binderVariancesAsymptoticallyNormal1983}. Building on this formulation, we derive a design-based sandwich variance estimator by combining the stacked GAMLSS score contributions, the corresponding Hessian (bread matrix), and an estimator of the survey-design variance (meat matrix) based on primary sampling unit (PSU)-level score contributions.

\subsection{Generalised additive models for location, scale, and shape}

Let $U_N=\{1,\ldots,N\}$ denote a finite population and let $\mathbf{Y}=(Y_1,\ldots,Y_N)^{\top}$ be a vector of response variable observations. Further, let $\mathcal{D}(\mu,\sigma,\nu,\tau)$ be a distribution with parameters $\mu$ (location), $\sigma$ (scale), and shape parameters $\nu$ and $\tau$, which are often associated with skewness and kurtosis, respectively:
\begin{align*}
    Y_i \sim \mathcal{D}(\mu_i,\sigma_i,\nu_i,\tau_i)
    \quad \text{for } i = 1,\ldots,N.
\end{align*}
\noindent In the present design-based setting, this notation is used to specify the working GAMLSS distribution for each finite-population unit. It should not be interpreted as specifying the joint distribution induced by the complex sampling design.

Let $m = 1,\ldots,M$ denote the number of distributional parameters included in the model. Further, let $g_m(\cdot)$ be known monotone link functions that connect the distributional parameters to a set of regressors in an additive way: 
\begin{equation}
\label{eqn:gamlss_parametric}
\begin{aligned}
g_1(\mu_i) &= \eta_{i1}
= \beta_{01} + \sum_{k=1}^{J_1} \beta_{k1} x_{ik}, \\
g_2(\sigma_i) &= \eta_{i2}
= \beta_{02} + \sum_{k=1}^{J_2} \beta_{k2} x_{ik}, \\
g_3(\nu_i) &= \eta_{i3}
= \beta_{03} + \sum_{k=1}^{J_3} \beta_{k3} x_{ik}, \\
g_4(\tau_i) &= \eta_{i4}
= \beta_{04} + \sum_{k=1}^{J_4} \beta_{k4} x_{ik}.
\end{aligned}
\end{equation}

\noindent For the derivation of the variance estimator, it is useful to rewrite Equation~(\ref{eqn:gamlss_parametric}) in compact unit-level notation. For unit $i\in U_N$, write $(\vartheta_{i1},\ldots,\vartheta_{iM})$ for the vector of unit-specific distribution parameters. 
\noindent Let $\mathbf{x}_{im}^{\top}$ denote all covariate values associated with the $i$-th observation unit. 
Then the parametric GAMLSS predictor for the $m$-th distributional parameter can be written as
\begin{align*}
    \eta_{im}
    =
    x_{im}^{\top}\boldsymbol{\beta}_m,
    \qquad
    g_m(\vartheta_{im})=\eta_{im},
    \qquad
    m=1,\ldots,M.
\end{align*}
Equivalently, $\vartheta_{im} = g_m^{-1}(x_{im}^{\top}\boldsymbol{\beta}_m)$. The observed data vector for population unit $i$ is denoted by $D_i = (y_i,x_{i1}^{\top},\ldots,x_{iM}^{\top})^{\top}$.
The coefficient vector for the $m$-th distributional parameter is $\boldsymbol{\beta}_m\in\mathbb{R}^{p_m}$. The full GAMLSS coefficient vector is obtained by stacking the parameter-specific coefficient vectors:
\begin{align*}
\btheta
=
\begin{pmatrix}
\bbeta_1\\
\bbeta_2\\
\vdots\\
\bbeta_M
\end{pmatrix}.
\end{align*}

\noindent Let $f_{\mathcal{D}}$ denote the density or probability mass function associated with the chosen GAMLSS family $\mathcal{D}$. The unit log-likelihood contribution is
\begin{align*}
    \ell_i(\boldsymbol{\theta})
    =
    \log f_{\mathcal{D}}
    \left\{
        y_i
        \mid
        \vartheta_{i1}(\boldsymbol{\beta}_1),
        \ldots,
        \vartheta_{iM}(\boldsymbol{\beta}_M)
    \right\}.
\end{align*}
The corresponding finite-population log-likelihood criterion is
\begin{align*}
    \ell_N(\boldsymbol{\theta})
    =
    \sum_{i\in U_N}\ell_i(\boldsymbol{\theta}).
\end{align*}
This criterion is used only to define the finite-population target. It is a census criterion over the fixed finite-population values and not a likelihood for the sample selection process.

\noindent The finite-population GAMLSS target $\boldsymbol{\theta}_N$ is defined as the solution of the census score equation $W_N^G(\boldsymbol{\theta}_N) = 0$, where
\begin{align*}
    W_N^G(\boldsymbol{\theta})
    =
    \sum_{i\in U_N}u_i^G(\boldsymbol{\theta}),
    \qquad
    u_i^G(\boldsymbol{\theta})
    =
    \frac{\partial \ell_i(\boldsymbol{\theta})}{\partial \boldsymbol{\theta}}.
\end{align*}
Thus, $\boldsymbol{\theta}_N$ is the coefficient vector that would be obtained by fitting the chosen parametric GAMLSS family and covariate specifications to the entire finite population. If the chosen GAMLSS family is only an approximation to the population distribution, $\boldsymbol{\theta}_N$ remains well-defined as a finite-population pseudo-maximum-likelihood parameter.

This places the parametric GAMLSS model in \citet{binderVariancesAsymptoticallyNormal1983}'s implicit finite-population parameter framework. \citet{binderVariancesAsymptoticallyNormal1983}'s general parameter equation can be written as
\begin{align*}
    W_N(\boldsymbol{\theta})
    =
    \sum_{i\in U_N}u(D_i;\boldsymbol{\theta})
    -
    v(\boldsymbol{\theta})
    =
    0.
\end{align*}
For the unpenalised parametric GAMLSS considered here, this framework is obtained by setting
\begin{align*}
    u(D_i;\boldsymbol{\theta})
    =
    u_i^G(\boldsymbol{\theta})
    =
    \frac{\partial \ell_i(\boldsymbol{\theta})}{\partial \boldsymbol{\theta}},
    \qquad
    v(\boldsymbol{\theta})=0.
\end{align*}
Consequently, the estimating contribution in Binder's framework corresponds to the full stacked GAMLSS score vector, which accounts for the joint estimation of all distributional parameters. Extensions incorporating fixed penalties would introduce an additional term through $v(\boldsymbol{\theta})$; however, the present derivation focuses on the unpenalised finite-dimensional parametric case.

\subsection{The stacked GAMLSS score}
\label{subsec:stacked_gamlss_score}

The estimating contribution used in the \citet{binderVariancesAsymptoticallyNormal1983}-type derivation is the unit GAMLSS score. Throughout, scores are written as column vectors. Since the full coefficient vector $\btheta$ is stacked by distributional parameter, the unit score is stacked in the same way:
\begin{align*}
\label{eq:unit-score-stacked}
    u_i^G(\btheta)
    =
    \frac{\partial \ell_i(\btheta)}{\partial \btheta}
    =
    \begin{pmatrix}
        u_{i1}^G(\btheta)\\
        u_{i2}^G(\btheta)\\
        \vdots\\
        u_{iM}^G(\btheta)
    \end{pmatrix},
    \qquad
    u_{im}^G(\btheta)
    =
    \frac{\partial \ell_i(\btheta)}{\partial \bbeta_m}.
\end{align*}

\noindent To derive one block of this vector, recall that $\eta_{im} =  x_{im}^{\trans}\bbeta_m$. Using the chain rule,
\begin{align*}
    u_{im}^G(\btheta)
    =
    \frac{\partial \ell_i(\btheta)}{\partial \bbeta_m}
    =
    \frac{\partial \ell_i(\btheta)}{\partial \eta_{im}}
    \frac{\partial \eta_{im}}{\partial \bbeta_m}.
\end{align*}
Since $\partial \eta_{im}/\partial \bbeta_m = x_{im}$, we obtain
\begin{align*}
    u_{im}^G(\btheta)
    =
    x_{im}q_{im}(\btheta),
    \qquad
    q_{im}(\btheta)
    =
    \frac{\partial \ell_i(\btheta)}{\partial \eta_{im}}.
\end{align*}
Here, $q_{im}(\btheta)$ is the scalar score contribution for the $m$-th distributional parameter on the predictor scale.
\noindent Equivalently, because $\vartheta_{im}  = g_m^{-1}(\eta_{im})$, another application of the chain rule gives
\begin{align*}
    q_{im}(\btheta)
    =
    \frac{\partial \ell_i(\btheta)}{\partial \vartheta_{im}}
    \frac{\partial \vartheta_{im}}{\partial \eta_{im}}.
\end{align*}
Defining
\begin{align*}
    \ell_{im}^{(1)}(\btheta)
    =
    \frac{\partial \ell_i(\btheta)}{\partial \vartheta_{im}},
\end{align*}
we can write $q_{im}(\btheta) = \ell_{im}^{(1)}(\btheta) (g_m^{-1})'(\eta_{im})$.
Therefore, $u_{im}^G(\btheta) = x_{im} \ell_{im}^{(1)}(\btheta) (g_m^{-1})'(\eta_{im})$. Combining the $M$ score blocks leads to the complete stacked GAMLSS score contribution, 
\begin{equation}
\label{eq:complete-stacked-score}
    u_i^G(\btheta)
    =
    \begin{pmatrix}
        x_{i1}q_{i1}(\btheta)\\
        x_{i2}q_{i2}(\btheta)\\
        \vdots\\
        x_{iM}q_{iM}(\btheta)
    \end{pmatrix}.
\end{equation}
This stacked score represents the GAMLSS analogue of the estimating contribution in \citet{binderVariancesAsymptoticallyNormal1983}'s framework. Whereas models with a single distributional parameter involve one score block, GAMLSS jointly estimates multiple distributional parameters. Consequently, all parameter-specific score blocks must be retained together, as they jointly define the finite-population target and its covariance structure.

\subsection{Survey-weighted sample estimating equation}
\label{subsec:sample_estimating_equation}

Let $s\subset U_N$ denote the realised sample. For the first-stage with-replacement cluster approximation, let $h=1,\ldots,H$ index strata, let $j=1,\ldots,C_h$ index the sampled primary sampling units (PSUs) in stratum $h$, and let $s_{hj}$ denote the set of sampled final units in primary sampling unit $j$ of stratum $h$. Thus,
\begin{align*}
    s
    =
    \bigcup_{h=1}^{H}
    \bigcup_{j=1}^{C_h}
    s_{hj}.
\end{align*}
Let $a_i>0$ denote the analysis weight attached to sampled unit $i$. In the Horvitz--Thompson case, $a_i=1/\pi_i$, where $\pi_i$ is the final-unit inclusion probability.

The sample analogue of the finite-population GAMLSS score total $W_N^G(\btheta)$ is obtained by replacing the census total with its survey-weighted sample version:
\begin{align}
\label{eq:weighted-sample-score-total}
    \widehat{W}_s^G(\btheta)
    =
    \sum_{i\in s} a_i u_i^G(\btheta).
\end{align}
The survey-weighted GAMLSS estimator $\widehat{\btheta}$ is then defined as the solution of $\widehat{W}_s^G(\widehat{\btheta}) = 0$. Using the stacked score representation from Equation~(\ref{eq:complete-stacked-score}), Equation~(\ref{eq:weighted-sample-score-total}) can be written as
\begin{align*}
    \widehat{W}_s^G(\btheta)
    =
    \sum_{i\in s}
    a_i
    \begin{pmatrix}
        x_{i1}q_{i1}(\btheta)\\
        x_{i2}q_{i2}(\btheta)\\
        \vdots\\
        x_{iM}q_{iM}(\btheta)
    \end{pmatrix}.
\end{align*}

\noindent If the analysis weights are inverse inclusion probabilities, then $\widehat{W}_s^G(\btheta)$ is the Horvitz--Thompson estimator of the finite-population score total $W_N^G(\btheta)$ for fixed $\btheta$. If calibrated or rescaled weights are used instead, the estimator and the variance estimator should be interpreted relative to the chosen weighted estimating equation. In all cases, the same weights must be used consistently in the estimating equation, in the derivative matrix, and in the survey-design variance estimator derived below.

\subsection{The GAMLSS bread matrix}
\label{subsec:gamlss_bread}

The derivative of the unit stacked score is the unit Hessian matrix
\begin{equation*}
\label{eq:unit-hessian}
    H_i(\btheta)
    =
    \frac{\partial u_i^G(\btheta)}{\partial \btheta^{\trans}}
    =
    \frac{\partial^2 \ell_i(\btheta)}
    {\partial \btheta \partial \btheta^{\trans}}.
\end{equation*}
The corresponding finite-population derivative matrix is
\begin{equation*}
\label{eq:finite-population-derivative}
    A_N^G(\btheta)
    =
    \frac{\partial W_N^G(\btheta)}{\partial \btheta^{\trans}}
    =
    \sum_{i\in U_N} H_i(\btheta).
\end{equation*}
Its survey-weighted sample analogue is
\begin{equation*}
\label{eq:sample-derivative}
    \widehat{A}_s^G(\btheta)
    =
    \frac{\partial \widehat{W}_s^G(\btheta)}
    {\partial \btheta^{\trans}}
    =
    \sum_{i\in s} a_i H_i(\btheta).
\end{equation*}

\noindent It is often convenient to write the final sandwich in terms of the observed information matrix rather than the derivative matrix. Define
\begin{equation}
\label{eq:observed-information}
    B_N^G(\btheta)
    =
    -A_N^G(\btheta),
    \qquad
    \widehat{B}_s^G(\btheta)
    =
    -\widehat{A}_s^G(\btheta).
\end{equation}
The sign convention does not affect the final covariance estimator because the bread appears on both sides of the meat.

\noindent Because $\btheta$ is stacked by distributional parameter, the Hessian $H_i(\btheta)$ is a block matrix. Its $(m,r)$ block is
\begin{equation*}
\label{eq:hessian-block-definition}
    H_{i,mr}(\btheta)
    =
    \frac{\partial u_{im}^G(\btheta)}
    {\partial \bbeta_r^{\trans}},
    \qquad
    m,r=1,\ldots,M.
\end{equation*}
Using $u_{im}^G(\btheta)=x_{im}q_{im}(\btheta)$ and treating the covariates as fixed finite-population quantities gives
\begin{equation}
\label{eq:hessian-block-link-scale}
    H_{i,mr}(\btheta)
    =
    x_{im}x_{ir}^{\trans}
    \frac{\partial^2 \ell_i(\btheta)}
    {\partial \eta_{im}\partial \eta_{ir}}.
\end{equation}
Thus, the full unit Hessian can be written schematically as
\begin{equation*}
\label{eq:hessian-block-matrix}
    H_i(\btheta)
    =
    \begin{pmatrix}
        H_{i,11}(\btheta) & H_{i,12}(\btheta) & \cdots & H_{i,1M}(\btheta)\\
        H_{i,21}(\btheta) & H_{i,22}(\btheta) & \cdots & H_{i,2M}(\btheta)\\
        \vdots & \vdots & \ddots & \vdots\\
        H_{i,M1}(\btheta) & H_{i,M2}(\btheta) & \cdots & H_{i,MM}(\btheta)
    \end{pmatrix}.
\end{equation*}

\noindent To express the link-scale curvature in terms of the distributional parameters, define
\begin{equation*}
\label{eq:second-parameter-derivative}
    \ell_{i,mr}^{(2)}(\btheta)
    =
    \frac{\partial^2 \ell_i(\btheta)}
    {\partial \vartheta_{im}\partial \vartheta_{ir}}.
\end{equation*}
Since $\vartheta_{im}=g_m^{-1}(\eta_{im})$, the chain rule gives
\begin{equation*}
\label{eq:link-scale-second-derivative}
\begin{aligned}
    \frac{\partial^2 \ell_i(\btheta)}
    {\partial \eta_{im}\partial \eta_{ir}}
    &=
    \ell_{i,mr}^{(2)}(\btheta)
    \left(g_m^{-1}\right)'(\eta_{im})
    \left(g_r^{-1}\right)'(\eta_{ir})\\
    &\qquad
    +
    \mathbbm{1}(m=r)
    \ell_{im}^{(1)}(\btheta)
    \left(g_m^{-1}\right)''(\eta_{im}).
\end{aligned}
\end{equation*}
Substitution into Equation~(\ref{eq:hessian-block-link-scale}) yields
\begin{equation*}
\label{eq:hessian-block-expanded}
\begin{aligned}
    H_{i,mr}(\btheta)
    &=
    x_{im}x_{ir}^{\trans}
    \biggl[
        \ell_{i,mr}^{(2)}(\btheta)
        \left(g_m^{-1}\right)'(\eta_{im})
        \left(g_r^{-1}\right)'(\eta_{ir})\\
    &\qquad\quad
        +
        \mathbbm{1}(m=r)
        \ell_{im}^{(1)}(\btheta)
        \left(g_m^{-1}\right)''(\eta_{im})
    \biggr].
\end{aligned}
\end{equation*}
For $m\neq r$, the second term disappears:
\begin{equation*}
\label{eq:hessian-off-diagonal-block}
    H_{i,mr}(\btheta)
    =
    x_{im}x_{ir}^{\trans}
    \ell_{i,mr}^{(2)}(\btheta)
    \left(g_m^{-1}\right)'(\eta_{im})
    \left(g_r^{-1}\right)'(\eta_{ir}).
\end{equation*}

\noindent The off-diagonal blocks are important. They show that the bread is not obtained by treating the distributional parameter equations as independent regressions. For example, the $\mu$-$\sigma$ block is generally
\begin{equation*}
\label{eq:mu-sigma-bread-block}
    H_{i,\mu\sigma}(\btheta)
    =
    x_{i\mu}x_{i\sigma}^{\trans}
    \frac{\partial^2 \ell_i(\btheta)}
    {\partial \eta_{i\mu}\partial \eta_{i\sigma}},
\end{equation*}
which need not be zero. The GAMLSS bread therefore contains the curvature of the full joint distributional model across all fitted parameters.

\subsection{Taylor linearisation of the implicit estimator}
\label{subsec:taylor_linearisation}

The estimator $\widehat{\btheta}$ is defined implicitly by $\widehat{W}_s^G(\widehat{\btheta}) = 0$.
To derive its design-based variance, expand $\widehat{W}_s^G(\widehat{\btheta})$ around the finite-population target $\btheta_N$. A first-order Taylor expansion gives
\begin{equation}
\label{eq:taylor-expansion}
    0
    =
    \widehat{W}_s^G(\widehat{\btheta})
    \approx
    \widehat{W}_s^G(\btheta_N)
    +
    \widehat{A}_s^G(\btheta_N)
    \left(
        \widehat{\btheta}
        -
        \btheta_N
    \right),
\end{equation}
where
\begin{equation*}
\label{eq:sample-derivative-repeat}
    \widehat{A}_s^G(\btheta)
    =
    \frac{\partial \widehat{W}_s^G(\btheta)}
    {\partial \btheta^{\trans}}.
\end{equation*}
Solving Equation~(\ref{eq:taylor-expansion}) for
$\widehat{\btheta}-\btheta_N$ gives the linearised approximation
\begin{equation}
\label{eq:linearised-estimator-sample-A}
    \widehat{\btheta}
    -
    \btheta_N
    \approx
    -
    \left\{
        \widehat{A}_s^G(\btheta_N)
    \right\}^{-1}
    \widehat{W}_s^G(\btheta_N).
\end{equation}

\noindent Since $\btheta_N$ is defined by the finite-population census score equation, $W_N^G(\btheta_N) = 0$, the random quantity $\widehat{W}_s^G(\btheta_N)$ can be interpreted as the survey estimation error of the finite-population score total at $\btheta_N$. More explicitly,
\begin{equation*}
\label{eq:score-total-error-variance}
    \Var_p
    \left\{
        \widehat{W}_s^G(\btheta_N)
    \right\}
    =
    \Var_p
    \left\{
        \widehat{W}_s^G(\btheta_N)
        -
        W_N^G(\btheta_N)
    \right\}.
\end{equation*}

Thus, the leading source of sampling variability is the design variance of the survey-weighted GAMLSS score total evaluated at the finite-population root.

For the asymptotic variance expression, the sample derivative matrix in Equation~(\ref{eq:linearised-estimator-sample-A}) is replaced by its finite-population counterpart
\begin{equation*}
\label{eq:finite-population-bread-repeat}
    A_N^G(\btheta_N)
    =
    \frac{\partial W_N^G(\btheta)}
    {\partial \btheta^{\trans}}
    \bigg|_{\btheta=\btheta_N}.
\end{equation*}
This yields $\widehat{\btheta}
    -
    \btheta_N
    \approx
    -
    \{
        A_N^G(\btheta_N)
    \}^{-1}
    \{
        \widehat{W}_s^G(\btheta_N)
        -
        W_N^G(\btheta_N)
    \}$. Taking the design variance on both sides gives
\begin{equation*}
\label{eq:linearised-design-variance-A}
    \Var_p(\widehat{\btheta})
    \approx
    \left\{
        A_N^G(\btheta_N)
    \right\}^{-1}
    \Omega_N^G(\btheta_N)
    \left\{
        A_N^G(\btheta_N)
    \right\}^{-\trans},
\end{equation*}
where
\begin{equation}
\label{eq:population-meat-definition}
    \Omega_N^G(\btheta_N)
    =
    \Var_p
    \left\{
        \widehat{W}_s^G(\btheta_N)
    \right\}
    =
    \Var_p
    \left\{
        \sum_{i\in s}a_i u_i^G(\btheta_N)
    \right\}.
\end{equation}
Here, $\Var_p(\cdot)$ denotes variance with respect to the sampling design, treating the finite-population values as fixed.

Using the observed information notation from Equation~(\ref{eq:observed-information}), the same approximation can be written as
\begin{equation}
\label{eq:linearised-design-variance-B}
    \Var_p(\widehat{\btheta})
    \approx
    \left\{
        B_N^G(\btheta_N)
    \right\}^{-1}
    \Omega_N^G(\btheta_N)
    \left\{
        B_N^G(\btheta_N)
    \right\}^{-\trans}.
\end{equation}
The sign change from $A_N^G$ to $B_N^G$ has no effect on the covariance matrix because the bread appears on both sides of the meat.

\subsection{One-stage clustered survey meat}
\label{subsec:survey_meat}

The matrix $\Omega_N^G(\btheta_N)$ in Equation~(\ref{eq:population-meat-definition}) is the design variance of the survey-weighted stacked score total. It is unknown because it depends on the full sampling design and on the unknown finite-population target $\btheta_N$. The one-stage plug-in estimator replaces $\btheta_N$ by the fitted value $\widehat{\btheta}$ and estimates the design variance from the sampled primary sampling unit score totals.

\noindent At the fitted value $\widehat{\btheta}$, define the weighted individual stacked score contribution as $\widehat{\psi}_i = a_i u_i^G(\widehat{\btheta})$. This is a $p\times 1$ vector. Using the block representation of the GAMLSS score, it can be written as
\begin{equation*}
\label{eq:weighted-individual-score-block}
    \widehat{\psi}_i
    =
    a_i
    \begin{pmatrix}
        x_{i1}\widehat{q}_{i1}\\
        x_{i2}\widehat{q}_{i2}\\
        \vdots\\
        x_{iM}\widehat{q}_{iM}
    \end{pmatrix},
    \qquad
    \widehat{q}_{im}
    =
    q_{im}(\widehat{\btheta}).
\end{equation*}

\noindent For sampled primary sampling unit $j$ in stratum $h$, define the PSU-level stacked score total as
\begin{equation*}
\label{eq:psu-score-total}
    \widehat{T}_{hj}
    =
    \sum_{i\in s_{hj}}\widehat{\psi}_i
    =
    \sum_{i\in s_{hj}}a_i u_i^G(\widehat{\btheta}).
\end{equation*}
This vector has the same dimension as $\btheta$. In block form,
\begin{equation*}
\label{eq:psu-score-total-block}
    \widehat{T}_{hj}
    =
    \begin{pmatrix}
        \widehat{T}_{hj,1}\\
        \widehat{T}_{hj,2}\\
        \vdots\\
        \widehat{T}_{hj,M}
    \end{pmatrix},
    \qquad
    \widehat{T}_{hj,m}
    =
    \sum_{i\in s_{hj}}a_i x_{im}\widehat{q}_{im}.
\end{equation*}
Thus, the score contributions are first evaluated at the observation level, then weighted, then aggregated to PSU-level totals.

The within-stratum mean PSU score total is
\begin{equation*}
\label{eq:mean-psu-score-total}
    \overline{T}_h
    =
    \frac{1}{C_h}
    \sum_{j=1}^{C_h}\widehat{T}_{hj},
\end{equation*}
where $C_h$ is the number of sampled PSUs per stratum $h$.


\noindent The one-stage stratified with-replacement estimator of the meat is then
\begin{equation}
\label{eq:wr-meat}
    \widehat{\Omega}_{\mathrm{WR}}^G
    =
    \sum_{h=1}^{H}
    \frac{C_h}{C_h-1}
    \sum_{j=1}^{C_h}
    \left(
        \widehat{T}_{hj}
        -
        \overline{T}_h
    \right)
    \left(
        \widehat{T}_{hj}
        -
        \overline{T}_h
    \right)^{\trans}.
\end{equation}
This formula requires $C_h\geq 2$ for every stratum used in variance estimation. If a stratum contains only one sampled PSU, it must be combined with another stratum or handled by an explicit lonely-PSU rule. The estimator in Equation~(\ref{eq:wr-meat}) is a with-replacement first-stage cluster approximation and therefore contains no finite population correction.

Stratification enters through centering PSU totals within strata. Clustering enters through forming cross-products after aggregation to PSU-level score totals. Unequal weighting enters through the weighted individual scores $\widehat{\psi}_i$.

The block structure of the meat follows directly from the block structure of the stacked score. Its $(m,r)$ block is
\begin{equation*}
\label{eq:wr-meat-block}
\begin{aligned}
    \widehat{\Omega}_{\mathrm{WR},mr}^G
    &=
    \sum_{h=1}^{H}
    \frac{C_h}{C_h-1}
    \sum_{j=1}^{C_h}
    \left(
        \widehat{T}_{hj,m}
        -
        \overline{T}_{h,m}
    \right)
    \left(
        \widehat{T}_{hj,r}
        -
        \overline{T}_{h,r}
    \right)^{\trans},
\end{aligned}
\end{equation*}

where $\overline T_{h,m}$ denotes the $m$-th block of $\overline T_h$ and $m,r=1,\ldots,M$.

\noindent For example, in a four-parameter model the $\mu$-$\sigma$ block is
\begin{equation*}
\label{eq:wr-meat-mu-sigma}
    \widehat{\Omega}_{\mathrm{WR},\mu\sigma}^G
    =
    \sum_{h=1}^{H}
    \frac{C_h}{C_h-1}
    \sum_{j=1}^{C_h}
    \left(
        \widehat{T}_{hj,\mu}
        -
        \overline{T}_{h,\mu}
    \right)
    \left(
        \widehat{T}_{hj,\sigma}
        -
        \overline{T}_{h,\sigma}
    \right)^{\trans}.
\end{equation*}
Hence, covariance between score blocks for different GAMLSS distributional parameters is estimated directly by the matrix cross-products of the stacked PSU totals. It is not added as a separate correction after fitting independent equations.

\subsection{Final one-stage survey-robust covariance estimator}
\label{subsec:final_covariance_estimator}

The final estimator is obtained by replacing the unknown finite-population quantities in Equation~(\ref{eq:linearised-design-variance-B}) with their sample plug-in analogues. The observed information matrix is evaluated at the fitted value $\widehat{\btheta}$,
\begin{equation*}
\label{eq:final-bread}
    \widehat{B}_s^G
    =
    \widehat{B}_s^G(\widehat{\btheta})
    =
    -
    \widehat{A}_s^G(\widehat{\btheta})
    =
    -
    \sum_{i\in s}a_i H_i(\widehat{\btheta}).
\end{equation*}
The meat is the one-stage stratified with-replacement design variance estimator of the weighted stacked score total:
\begin{equation*}
\label{eq:final-meat}
    \widehat{\Omega}_{\mathrm{WR}}^G
    =
    \sum_{h=1}^{H}
    \frac{C_h}{C_h-1}
    \sum_{j=1}^{C_h}
    \left(
        \widehat{T}_{hj}
        -
        \overline{T}_h
    \right)
    \left(
        \widehat{T}_{hj}
        -
        \overline{T}_h
    \right)^{\trans}.
\end{equation*}
The resulting one-stage survey-robust covariance estimator for the estimated GAMLSS coefficient vector is therefore
\begin{equation*}
\label{eq:final-sandwich}
    \widehat{\operatorname{Var}}_p(\widehat{\btheta})
    =
    \left(
        \widehat{B}_s^G
    \right)^{-1}
    \widehat{\Omega}_{\mathrm{WR}}^G
    \left(
        \widehat{B}_s^G
    \right)^{-\trans}.
\end{equation*}

\noindent The corresponding coefficient-level standard error for the $k$-th element of $\widehat{\btheta}$ is
\begin{equation*}
\widehat{\operatorname{se}}(\widehat{\theta}_k)
=
\left\{
    \left[
        \widehat{\operatorname{Var}}_p(\widehat{\btheta})
    \right]_{kk}
\right\}^{1/2}.
\end{equation*}


\section{Simulation experiments}

\subsection{Simulation design}

We conducted a simulation study with 10 scenarios to assess the calibration of standard errors for parametric GAMLSS fits under increasingly complex conditions. The scenarios varied two sources of complexity: (i) the outcome model, through covariate effects on location, scale, and shape parameters, and (ii) the sampling design, through stratification, clustering, unequal weights, PPS sampling, and limited numbers of sampled PSUs. Scenarios S1--S4 serve as benchmarks under simple random sampling, whereas S5--S10 represent increasingly realistic complex household survey designs. Table~\ref{tab:scenario_overview} provides an overview of the scenarios. The following subsections describe the finite population, data-generating mechanisms, Monte Carlo procedure, and evaluation criteria.

\begin{table*}[t]
\caption{Overview of the simulation scenarios. All scenarios use $n=24{,}395$ sampled households per Monte Carlo repetition.\label{tab:scenario_overview}}
\centering
\small
\tabcolsep=3pt
\renewcommand{\arraystretch}{1.15}

\begin{tabularx}{\textwidth}{@{}c l l Y Y Y@{}}
\toprule
\textbf{Scenario} &
\textbf{Family} &
\textbf{Active pars.} &
\textbf{DGP population structure} &
\textbf{Sampling design / weights} &
\textbf{Purpose} \\
\midrule

\textbf{S1} &
Normal &
$\mu$ &
None &
SRS; no clustering or stratification; constant weights &
Calibration benchmark \\

\textbf{S2} &
Normal &
$\mu,\sigma$ &
None &
SRS; no clustering or stratification; constant weights &
Isolate heteroskedasticity beyond mean regression \\

\textbf{S3} &
BCPEo &
$\mu,\sigma,\nu$ &
None &
SRS; no clustering or stratification; constant weights &
Assess covariate-dependent shape/skewness under simple sampling \\

\textbf{S4} &
BCPEo &
$\mu,\sigma,\nu,\tau$ &
None &
SRS; no clustering or stratification; constant weights &
Full distributional-regression benchmark under simple sampling \\

\addlinespace[0.5em]

\textbf{S5} &
BCPEo &
$\mu,\sigma,\nu,\tau$ &
Stratum-aligned effects in $\mu$ and $\sigma$; strata are outcome-relevant &
Stratified SRS within 62 region $\times$ urban/rural strata; near-constant weights, mean weight CV $\approx 0.002$ &
Check behaviour when stratification is aligned with outcome structure \\

\textbf{S6} &
BCPEo &
$\mu,\sigma,\nu,\tau$ &
Cluster effect in $\mu$ with target SD $0.10$ on the linear-predictor scale; empirical outcome ICC $\approx 0.264$ &
Two-stage cluster sampling; 400 sampled PSUs; one analysis stratum; mean weight CV $\approx 0.677$ &
Test moderate outcome clustering under two-stage sampling \\

\textbf{S7} &
BCPEo &
$\mu,\sigma,\nu,\tau$ &
Stronger cluster effects in $\mu$ and $\sigma$, with target SDs $0.28$ and $0.16$ on the linear-predictor scale; empirical outcome ICC $\approx 0.479$ &
Two-stage cluster sampling; 250 sampled PSUs; one analysis stratum; mean weight CV $\approx 0.449$ &
Test strong outcome clustering and distributional heterogeneity \\

\textbf{S8} &
BCPEo &
$\mu,\sigma,\nu,\tau$ &
Cluster effects and cluster-size effects in $\mu$ and $\sigma$; empirical outcome ICC $\approx 0.532$ &
PPS / unequal-probability multistage sampling; 250 sampled PSUs; one analysis stratum; mean weight CV $\approx 0.206$ &
Quantify PPS and cluster-size-linked outcome structure \\

\textbf{S9} &
BCPEo &
$\mu,\sigma,\nu,\tau$ &
Same population DGP as S7 &
Two-stage cluster sampling with fewer PSUs; 150 sampled PSUs; one analysis stratum; mean weight CV $\approx 0.055$ &
Few-PSU stress test under the S7 population DGP \\

\textbf{S10} &
BCPEo &
$\mu,\sigma,\nu,\tau$ &
Same population DGP as S8 &
PPS / unequal-probability multistage sampling with severe weight dispersion; 250 sampled PSUs; one analysis stratum; mean weight CV $\approx 0.932$ &
Severe-weight stress test \\

\botrule
\end{tabularx}

\begin{tablenotes}
\item Notes: SDs for stratum, cluster, and cluster-size effects are measured on the GAMLSS linear-predictor scale and do not represent target ICCs. Empirical ICCs are calculated from the generated finite population using a one-way cluster decomposition. Weight CV denotes the coefficient of variation of normalised design weights.
\end{tablenotes}
\end{table*}

\subsection{Synthetic finite population}

We generated an empirically calibrated synthetic finite population reflecting the multilevel structure of complex household surveys. The generative model was fitted to household-level data from the 2015 Afghanistan Demographic and Health Survey (DHS; \citealp{centralstatisticsorganizationcsoAfghanistanDemographicHealth2017}). After data cleaning and FAMD-based variable selection (\citealp{pagesAnalyseFactorielleDonnees2004}), 31 household-level variables were modelled using variable-specific marginal models with cluster random effects and stratum fixed effects. Dependence between variables at the cluster level was captured through a Gaussian copula.

The fitted marginal distributions and copula were used to generate the population cluster by cluster. Clusters were assigned to strata following the DHS design, while cluster sizes were generated from a proxy lognormal model based on observed cluster-level household counts. The resulting finite population contained approximately $4.1\times10^6$ households across $25{,}079$ clusters.

Following the ADEMP framework (\citealp{morrisUsingSimulationStudies2019}), this population served as a realistic test bed for evaluating variance estimation under complex survey designs. We assessed whether the proposed linearisation-based survey-robust sandwich estimator provides calibrated standard errors for parametric GAMLSS fits under clustering, stratification, and unequal weighting, while retaining efficiency under simpler designs.

We compared five variance estimators: the conventional model-based estimator, the model-robust GAMLSS estimator, bootstrap and delete-a-group Jackknife replication estimators (\citealp{efronBootstrapMethodsAnother1979,kottUsingDeleteaGroupJackknife2001}), and the proposed estimator. The target quantity was the repeated-sampling variance of GAMLSS parameter estimates under the specified population and sampling design.

\subsection{Outcome-generating models}

For each scenario, outcomes were generated once for all households in the synthetic finite population and kept fixed across Monte Carlo repetitions. Thus, the simulation variability arises exclusively from repeated sampling and model fitting. Scenarios S1--S2 used Gaussian outcomes, whereas S3--S10 used the Box--Cox Power Exponential distribution with original log-link for the location parameter (BCPEo; \citealp{rigbySmoothCentileCurves2004}). The BCPEo distribution allows all four distributional parameters to vary with covariates and provides flexible control over skewness and kurtosis.

Two nested sets of covariates were selected from the synthetic population using outcome-free screening. The smaller set was used for the Gaussian benchmarks and the first BCPEo scenario, while the larger set introduced additional distributional-regression complexity in the remaining scenarios (see Table~\ref{tab:term_counts} in the Appendix). Covariates were standardised and combined into parameter-specific linear predictors with controlled effect sizes.

For the Gaussian scenarios,
\[
Y_i\sim N(\mu_i,\sigma_i^2),
\]
with $\mu_i=100+\eta_{\mu i}$. Scenario S1 assumed constant variance ($\sigma_i=15$), whereas S2 allowed heteroskedasticity through $\sigma_i=15\exp(\eta_{\sigma i})$.

For the BCPEo scenarios,
\[
Y_i\sim\mathrm{BCPEo}(\mu_i,\sigma_i,\nu_i,\tau_i),
\]
with $\mu_i=100\exp(\eta_{\mu i}), \sigma_i=0.12\exp(\eta_{\sigma i}), \nu_i=\eta_{\nu i}$, and $\tau_i=2.5\exp(\eta_{\tau i})$.
Scenario S3 used the smaller covariate set with covariate effects on $\mu$, $\sigma$, and $\nu$, while keeping $\tau$ constant. From S4 onwards, the larger covariate set was used and all four distributional parameters depended on covariates.

To introduce realistic dependencies between outcomes and survey design features, additional population effects were added in the complex-survey scenarios. Stratum effects were introduced for $\mu$ and $\sigma$ in S5, cluster effects were added to $\mu$ in S6 and to both $\mu$ and $\sigma$ in S7 and S9, and cluster-size effects were additionally included in S8 and S10 to link outcome distributions with unequal-probability sampling mechanisms.

\subsection{Monte Carlo sampling and evaluation criteria}

The 10 scenarios form a structured progression along two dimensions: distributional-regression complexity and sampling-design complexity. Scenarios S1--S4 use simple random sampling and therefore isolate the effect of the fitted outcome model. S1 and S2 are Gaussian calibration benchmarks: S1 includes covariate effects only in the location parameter, while S2 additionally introduces covariate-dependent heteroskedasticity. Scenarios S3 and S4 replace the Gaussian outcome model by a BCPEo GAMLSS specification. S3 uses the moderate covariate set with covariate-dependent $\mu$, $\sigma$, and $\nu$, while keeping $\tau$ intercept-only; S4 uses the rich covariate set and allows all four BCPEo parameters to depend on covariates.

Scenarios S5--S10 keep the full BCPEo specification fixed and instead vary the finite-population and sampling-design features. S5 introduces outcome-relevant stratification by adding stratum-aligned components to the linear predictors for $\mu$ and $\sigma$. S6 and S7 introduce two-stage cluster sampling with increasing within-cluster dependence, first through a low-variance cluster effect in $\eta_\mu$ and then through stronger cluster effects in both $\eta_\mu$ and $\eta_\sigma$. S8 adds cluster-size-related outcome structure and unequal-probability multistage sampling. Finally, S9 and S10 are stress tests for the large-sample approximations underlying sandwich-type variance estimation: S9 uses the same population DGP as S7 but reduces the number of sampled primary sampling units, whereas S10 uses the same population DGP as S8 but imposes severe weight dispersion.

Table~\ref{tab:scenario_overview} summarizes the 10 scenarios. The reported numbers of terms refer to covariate effects in addition to intercepts. A dash indicates that the corresponding parameter is not part of the outcome distribution, while a zero indicates an intercept-only parameter.


For each scenario, $R=1000$ Monte Carlo samples were drawn from the fixed synthetic finite population. The sample size was set to the cleaned empirical DHS sample size, resulting in $n=24{,}395$ observations per repetition. In S1--S4, identical samples were used across scenario-specific outcomes within each repetition to reduce Monte Carlo noise when comparing baseline settings. For S5--S10, the complete survey design information, including sampled PSUs, strata, and design weights, was retained for each repetition.

The fitted GAMLSS models matched the respective data-generating mechanisms. Gaussian scenarios used the normal GAMLSS family with identity and log links for $\mu$ and $\sigma$, respectively. BCPEo models used log links for $\mu$, $\sigma$, and $\tau$, and an identity link for $\nu$. Models in S1--S4 were fitted without survey weights, whereas models in S5--S10 used the normalised design weights induced by the corresponding sampling designs. Convergence and estimation results were recorded for all repetitions.

Estimator performance was evaluated for each coefficient using the empirical Monte Carlo standard deviation of the fitted estimates as the reference repeated-sampling standard error. For variance estimator $m$, we computed the standard-error ratio
\begin{align*}
\mathrm{SER}_{j}^{(m)}
=
\frac{
\left\{R^{-1}\sum_{r=1}^{R}
\left(\widehat{\mathrm{SE}}_{rj}^{(m)}\right)^2
\right\}^{1/2}
}{
\operatorname{sd}\left(\hat{\theta}_{1j},\ldots,\hat{\theta}_{Rj}\right)
}.
\end{align*}
Values close to one indicate well-calibrated standard errors, values below one indicate underestimation, and values above one indicate conservative uncertainty estimates. For each scenario and distributional parameter, we report the median SER across active non-intercept coefficients. We additionally report centred coverage probabilities of nominal 95\% Wald intervals, using the Monte Carlo mean of each coefficient estimate as the reference value.

\subsection{Simulation results}

\subsubsection{Standard-error calibration}

\begin{table*}[t]
\caption{Median estimated-to-empirical standard-error ratios by GAMLSS parameter. Values close to one indicate good calibration. Dashes indicate parameters that are absent or intercept-only in the corresponding scenario. Detailed results can be found in the Appendix in Figures \ref{fig:appendix_se_ratio_boxplots_1}, \ref{fig:appendix_se_ratio_boxplots_2} and \ref{fig:appendix_se_ratio_boxplots_3}.}
\label{tab:SE_ratios_by_parameter}
\centering
\small
\setlength{\tabcolsep}{3pt}
\renewcommand{\arraystretch}{1.06}

\begin{tabular}{@{}cc@{}}
\begin{minipage}[t]{0.48\textwidth}
\centering
\textbf{(a) $\mu$ coefficients}

\vspace{0.25em}

\begin{tabular*}{\linewidth}{@{\extracolsep{\fill}}cccccG@{}}
\toprule
Scenario & Naive & Model-Robust & Bootstrap & Jackknife & \multicolumn{1}{c}{Survey-Robust}\\
\midrule
S1 & 0.99 & 0.99 & 0.98 & 0.98 & 0.99\\
S2 & 0.99 & 0.99 & 0.99 & 0.99 & 0.99\\
S3 & 0.99 & 1.00 & 0.99 & 1.00 & 1.00\\
S4 & 0.98 & 1.00 & 1.00 & 1.00 & 1.00\\
\addlinespace[0.5em]
S5 & 0.99 & 1.03 & 1.00 & 1.01 & 1.00\\
S6 & 0.34 & 0.36 & 0.98 & 1.03 & 0.98\\
S7 & 0.20 & 0.22 & 0.95 & 1.01 & 0.97\\
S8 & 0.25 & 0.26 & 0.97 & 1.02 & 0.99\\
S9 & 0.17 & 0.18 & 0.92 & 1.05 & 0.95\\
S10 & 0.19 & 0.20 & 0.98 & 1.05 & 0.97\\
\botrule
\end{tabular*}
\end{minipage}
&
\begin{minipage}[t]{0.48\textwidth}
\centering
\textbf{(b) $\sigma$ coefficients}

\vspace{0.25em}

\begin{tabular*}{\linewidth}{@{\extracolsep{\fill}}cccccG@{}}
\toprule
Scenario & Naive & Model-Robust & Bootstrap & Jackknife & \multicolumn{1}{c}{Survey-Robust}\\
\midrule
S1 & -- & -- & -- & -- & --\\
S2 & 1.01 & 1.01 & 1.02 & 1.03 & 1.01\\
S3 & 0.99 & 0.99 & 0.98 & 1.00 & 0.99\\
S4 & 0.99 & 0.99 & 1.00 & 1.00 & 0.99\\
\addlinespace[0.5em]
S5 & 1.01 & 1.04 & 1.00 & 1.00 & 1.00\\
S6 & 0.45 & 0.45 & 0.94 & 1.02 & 0.96\\
S7 & 0.21 & 0.21 & 0.92 & 1.03 & 0.92\\
S8 & 0.22 & 0.22 & 0.94 & 1.03 & 0.92\\
S9 & 0.16 & 0.17 & 0.90 & 1.03 & 0.90\\
S10 & 0.18 & 0.18 & 0.91 & 1.06 & 0.87\\
\botrule
\end{tabular*}
\end{minipage}
\\[1.0em]
\begin{minipage}[t]{0.48\textwidth}
\centering
\textbf{(c) $\nu$ coefficients}

\vspace{0.25em}

\begin{tabular*}{\linewidth}{@{\extracolsep{\fill}}cccccG@{}}
\toprule
Scenario & Naive & Model-Robust & Bootstrap & Jackknife & \multicolumn{1}{c}{Survey-Robust}\\
\midrule
S1 & -- & -- & -- & -- & --\\
S2 & -- & -- & -- & -- & --\\
S3 & 0.89 & 1.04 & 1.00 & 1.02 & 1.03\\
S4 & 0.89 & 0.96 & 0.98 & 1.00 & 0.96\\
\addlinespace[0.5em]
S5 & 1.01 & 1.02 & 1.01 & 1.02 & 1.00\\
S6 & 0.48 & 0.47 & 0.95 & 1.02 & 0.94\\
S7 & 0.21 & 0.21 & 0.93 & 1.05 & 0.88\\
S8 & 0.23 & 0.22 & 0.90 & 1.03 & 0.85\\
S9 & 0.18 & 0.18 & 0.93 & 1.06 & 0.88\\
S10 & 0.17 & 0.17 & 0.93 & 1.06 & 0.85\\
\botrule
\end{tabular*}
\end{minipage}
&
\begin{minipage}[t]{0.48\textwidth}
\centering
\textbf{(d) $\tau$ coefficients}

\vspace{0.25em}

\begin{tabular*}{\linewidth}{@{\extracolsep{\fill}}cccccG@{}}
\toprule
Scenario & Naive & Model-Robust & Bootstrap & Jackknife & \multicolumn{1}{c}{Survey-Robust}\\
\midrule
S1 & -- & -- & -- & -- & --\\
S2 & -- & -- & -- & -- & --\\
S3 & -- & -- & -- & -- & --\\
S4 & 0.98 & 0.99 & 1.00 & 1.00 & 0.99\\
\addlinespace[0.5em]
S5 & 0.90 & 1.01 & 0.99 & 1.01 & 1.00\\
S6 & 0.71 & 0.72 & 0.98 & 1.03 & 0.97\\
S7 & 0.26 & 0.27 & 0.97 & 1.06 & 0.96\\
S8 & 0.27 & 0.28 & 0.97 & 1.05 & 0.94\\
S9 & 0.22 & 0.24 & 1.02 & 1.13 & 1.04\\
S10 & 0.24 & 0.25 & 1.04 & 1.13 & 1.04\\
\botrule
\end{tabular*}
\end{minipage}
\end{tabular}
\end{table*}

\begin{figure*}[h]
\centering
\includegraphics[
  width=\textwidth,
  height=0.82\textheight,
  keepaspectratio
]{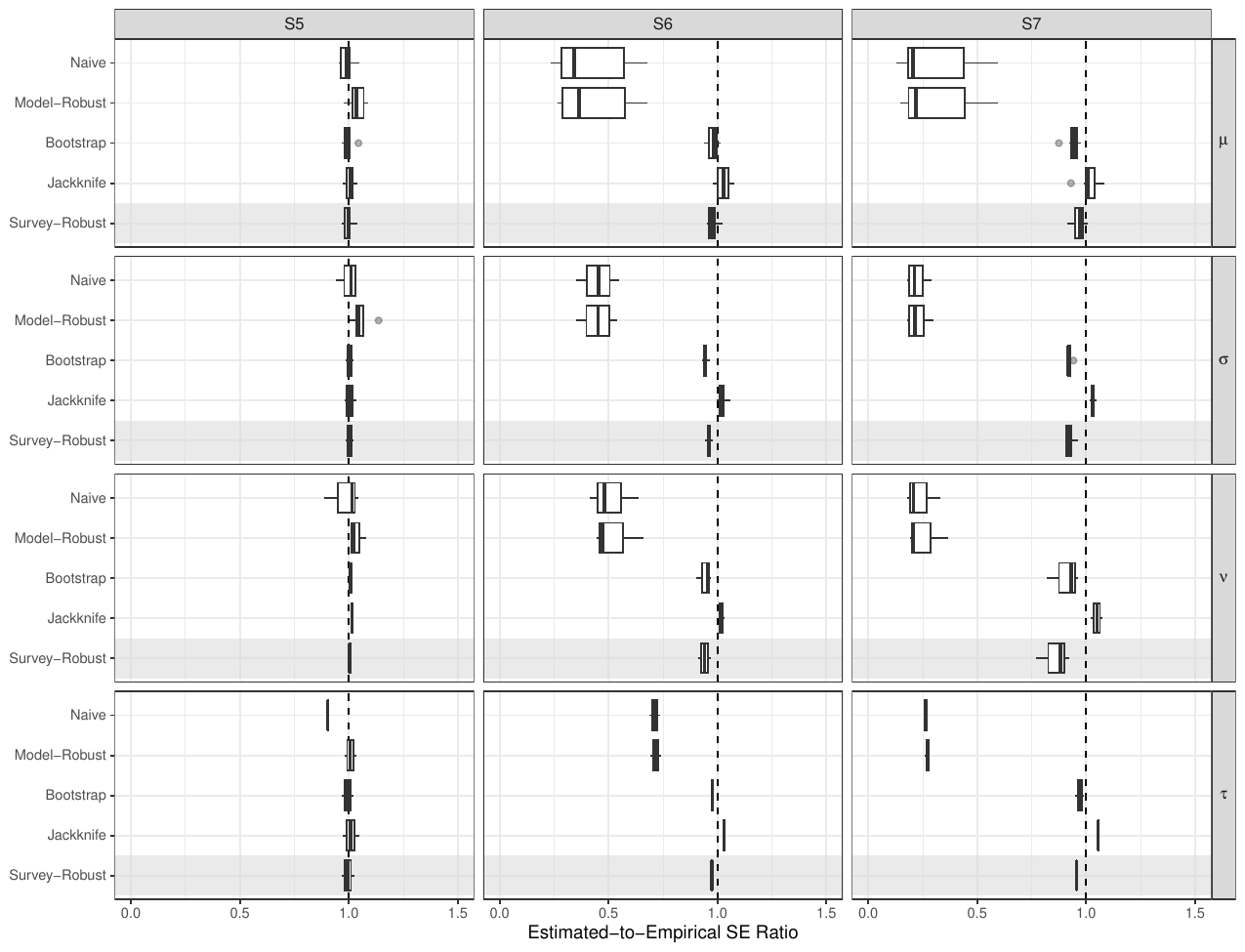}
\caption{Coefficient-level distributions of estimated-to-empirical standard-error ratios for scenarios S5--S7. Boxplots are shown separately by GAMLSS parameter, scenario, and variance estimator. Shaded in grey is the proposed survey-robust sandwich estimator. Intercepts are excluded. The dashed vertical reference line indicates perfect calibration. Values below one indicate underestimation of standard errors, while values above one indicate conservative standard errors.}
\label{fig:appendix_se_ratio_boxplots_2}
\end{figure*}

\begin{figure*}[h]
\centering
\includegraphics[
  width=\textwidth,
  height=0.82\textheight,
  keepaspectratio
]{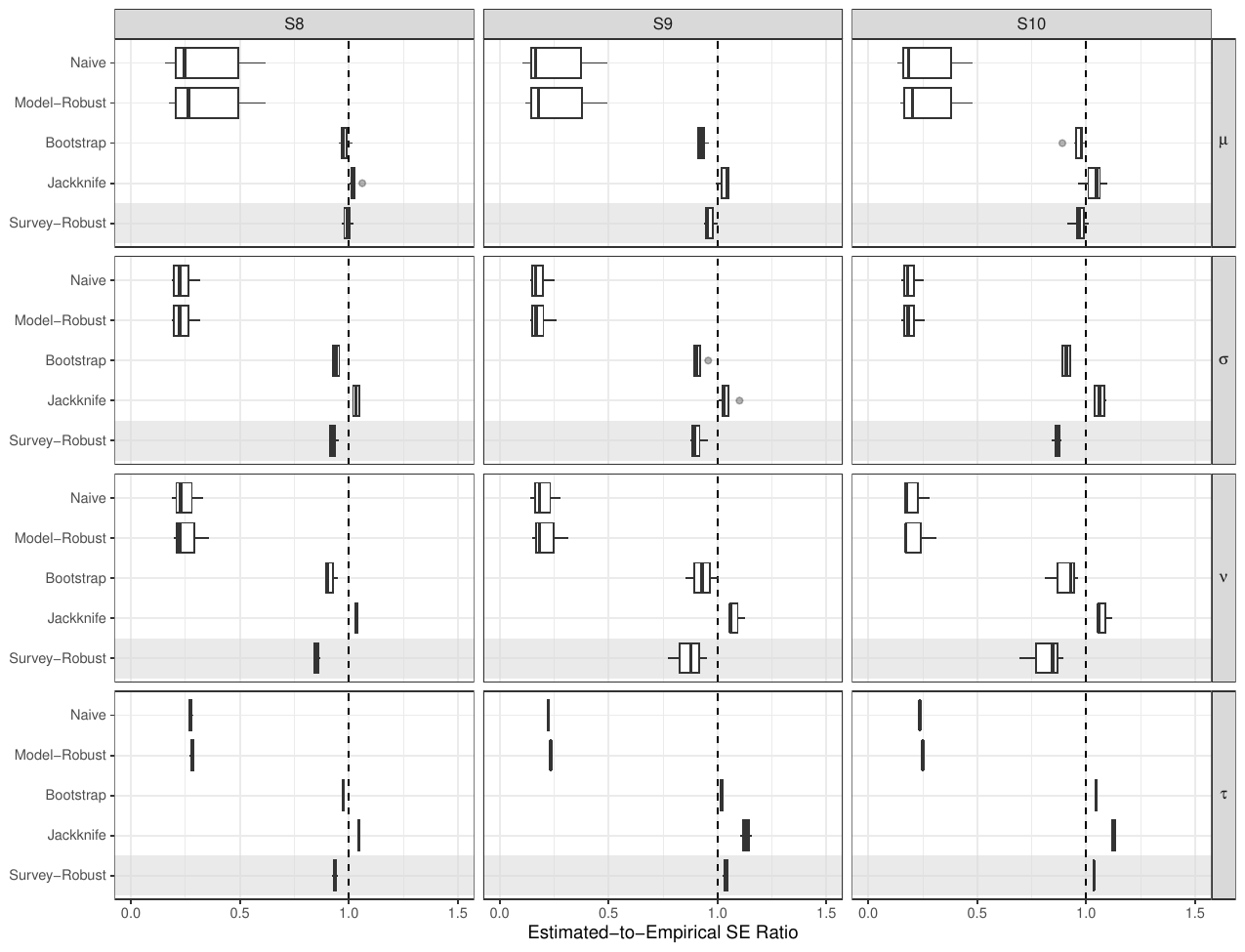}
\caption{Coefficient-level distributions of estimated-to-empirical standard-error ratios for scenarios S8--S10. Boxplots are shown separately by GAMLSS parameter, scenario, and variance estimator. Shaded in grey is the proposed survey-robust sandwich estimator. Intercepts are excluded. The dashed vertical reference line indicates perfect calibration. Values below one indicate underestimation of standard errors, while values above one indicate conservative standard errors.}
\label{fig:appendix_se_ratio_boxplots_3}
\end{figure*}

All scenario-specific GAMLSS models converged in every Monte Carlo repetition, and standard errors were obtained for all variance estimators considered. Table~\ref{tab:SE_ratios_by_parameter} summarises the median estimated-to-empirical standard-error ratios for the four distributional parameters; additional coefficient-level results are provided in the Appendix (Figures~\ref{fig:appendix_se_ratio_boxplots_1}--\ref{fig:appendix_se_ratio_boxplots_3}). Ratios close to one indicate well-calibrated standard errors, values below one indicate underestimation, and values above one indicate conservative uncertainty estimates.

Under simple random sampling (S1--S4), all estimators were well calibrated, with median ratios close to one across distributional parameters. For example, the ratios for $\mu$ ranged from 0.98 to 1.00, and those for $\sigma$ from 0.98 to 1.03. These results confirm that the proposed survey-robust estimator does not introduce unnecessary conservatism under near-independent sampling. Minor underestimation was observed for the naive estimator for $\nu$ in S3, whereas the proposed and replication-based estimators remained close to the empirical benchmark.

The stratified design scenario S5 showed similarly good calibration. This reflects the near-constant weights and outcome-relevant stratification structure, for which all estimators performed well. The proposed estimator achieved ratios close to one for all distributional parameters.

Substantial differences emerged once clustering was introduced. In S6 and S7, the naive and model-robust GAMLSS estimators severely underestimated uncertainty, with median ratios for $\mu$ decreasing to approximately 0.34 and 0.20 in S6 and S7, respectively. Similar patterns were observed for the remaining distributional parameters. In contrast, the proposed estimator remained well calibrated, with ratios between 0.88 and 0.97 in S7. The bootstrap and Jackknife estimators showed comparable accuracy, although the Jackknife was occasionally conservative.

The most challenging settings (S8--S10), which combined clustering with unequal-probability sampling and increased weight variability, reinforced these findings. Naive and model-robust estimators continued to substantially underestimate variability, whereas replication-based and survey-robust estimators remained close to the empirical benchmark. The proposed estimator achieved ratios between 0.85 and 1.04 across active parameters, while replication methods showed some conservatism and instability in the most demanding scenarios.

\subsubsection{Wald interval coverage}

\begin{table}[t]
\small
\caption{Median centred coverage of nominal 95\% Wald intervals across active non-intercept coefficients. Values are percentages.}
\label{tab:wald_centered_coverage}
\centering
\begin{tabular*}{\columnwidth}{@{\extracolsep{\fill}}cccccG@{}}
\toprule
Scenario & Naive & Model-Robust & Bootstrap & Jackknife & \multicolumn{1}{c}{Survey-Robust}\\
\midrule
S1 & 94.8 & 94.8 & 93.8 & 95.0 & 94.8\\
S2 & 95.0 & 95.0 & 94.5 & 94.5 & 95.0\\
S3 & 94.9 & 95.0 & 95.5 & 95.0 & 95.0\\
S4 & 94.6 & 94.7 & 95.0 & 95.0 & 94.7\\
\addlinespace[0.5em]
S5 & 94.4 & 95.8 & 95.0 & 94.5 & 94.7\\
S6 & 65.9 & 65.0 & 93.5 & 95.0 & 94.2\\
S7 & 34.4 & 35.3 & 92.0 & 95.0 & 93.3\\
S8 & 37.5 & 39.5 & 93.0 & 94.5 & 93.4\\
S9 & 28.0 & 27.9 & 92.5 & 95.0 & 93.3\\
S10 & 28.2 & 28.9 & 93.0 & 95.5 & 93.2\\
\botrule
\end{tabular*}
\end{table}

\begin{table}[h]
\caption{Post-fit computational time (seconds) for variance estimation. Replication times are medians over 10 sampled datasets per scenario using 100 bootstrap (B) and delete-a-group Jackknife (JK) refits. Survey-robust times are medians over 100 datasets and exclude model fitting. Ratios show replication-to-survey-robust time increases.}
\label{tab:variance_estimator_timing}
\centering
\small
\begin{tabular*}{\columnwidth}{@{\extracolsep{\fill}}cccGcc@{}}
\toprule
Scenario & Bootstrap & Jackknife & \multicolumn{1}{c}{Survey-robust} & Ratio (B) & Ratio (JK) \\
\midrule
S1  & 3.46   & 5.00   & 0.25 & 13.62 & 19.69 \\
S2  & 5.38   & 7.57   & 0.37 & 14.51 & 20.41 \\
S3  & 170.97 & 264.73 & 10.9 & 15.69 & 24.29 \\
S4  & 183.96 & 279.20 & 21.9 & 8.40 & 12.75 \\
\addlinespace[0.5em]
S5  & 309.81 & 486.27 & 22.0 & 14.08 & 22.10 \\
S6  & 179.77 & 278.94 & 21.2 & 8.48 & 13.16 \\
S7  & 248.88 & 349.73 & 21.3 & 11.68 & 16.42 \\
S8  & 291.45 & 415.89 & 21.3 & 13.68 & 19.53 \\
S9  & 312.86 & 430.86 & 21.4 & 14.62 & 20.13 \\
S10 & 306.27 & 423.42 & 24.3 & 12.60 & 17.42 \\
\botrule
\end{tabular*}
\end{table}

Table~\ref{tab:wald_centered_coverage} reports median centred coverage of nominal 95\% Wald intervals across active non-intercept coefficients. Since intervals are centred at the Monte Carlo mean, these results assess the calibration of the estimated standard errors relative to repeated-sampling variability rather than coefficient bias.

Under simple random sampling (S1--S4), all estimators achieved coverage close to the nominal level. The proposed survey-robust estimator yielded median coverage between 94.7\% and 95.0\%, confirming that it does not introduce unnecessary conservatism under near-independent sampling. Scenario S5 showed similarly good performance, reflecting the near-constant weights and absence of clustering.

In contrast, coverage of the naive and model-robust GAMLSS estimators deteriorated substantially once clustering and unequal weighting were introduced. In S6, median coverage decreased to approximately 65\%, and in the more demanding scenarios S7--S10 it fell below 40\% in some cases. This undercoverage directly reflects the corresponding underestimation of standard errors observed in Table~\ref{tab:SE_ratios_by_parameter}.

The proposed survey-robust estimator maintained near-nominal coverage throughout the complex-survey scenarios, with median coverage between 93.2\% and 94.7\% in S6--S10. Slight undercoverage occurred in the most challenging settings with fewer PSUs or greater weight variability, but performance remained substantially closer to the target level than the naive and model-robust estimators. Replication-based methods showed comparable coverage, with bootstrap occasionally slightly below nominal and the Jackknife somewhat conservative in several scenarios.

\subsubsection{Computational performance}

Table~\ref{tab:variance_estimator_timing} summarises the computational effort required for the different variance estimators. All calculations were performed on an Intel Core Ultra 9 285H processor with 96 GB RAM.

The proposed survey-robust estimator requires only a single model fit and computation of the corresponding variance components. Across scenarios, computation of the full variance-covariance matrix required less than one second in the simplest settings (S1--S2), approximately 11 seconds in S3, and around 21--24 seconds in the remaining scenarios.

In contrast, the bootstrap and Jackknife estimators required 100 additional GAMLSS refits per Monte Carlo repetition. Consequently, computational costs increased substantially with model complexity. While replication-based methods were only moderately more expensive in the simplest scenarios, they required several minutes per repetition in the more complex BCPEo settings. Overall, the bootstrap increased computational time by approximately 8--16-fold, whereas the Jackknife increased it by approximately 13--24-fold compared with the proposed estimator. Since practical applications often require 100--200 bootstrap repetitions (\citealp[p.~52]{efronIntroductionBootstrap1993}), the computational burden may be even greater.

These results demonstrate that the proposed estimator provides a computationally efficient alternative to replication-based variance estimation. It achieves comparable variance calibration while avoiding the repeated model fitting required by replication methods.

\section{Real-data application to childhood stunting determinants in Rwanda}

To illustrate the practical implications of the proposed estimator, we revisited the analysis of childhood stunting determinants in Rwanda by \citet{meurerDeterminantsChildhoodMalnutrition2024}. The analysis uses data from the 2019/20 DHS (\citealp{nationalinstituteofstatisticsofrwandanisrRwandaDemographicHealth2021}) and considers the height-for-age z-score of children under five years, a standardised measure relative to the WHO child growth reference curves (\citealp{worldhealthorganizationWHOChildGrowth2006}). Values below $-2$ and $-3$ indicate moderate and severe chronic undernutrition, respectively.

\citet{meurerDeterminantsChildhoodMalnutrition2024} modelled the z-score using a Johnson's $S_u$-distribution (JSUo; \citealp{johnsonSystemsFrequencyCurves1949}) within the GAMLSS framework:
\begin{align*}
    \texttt{HW70}_i \sim \mathrm{JSUo}(\mu_i,\sigma_i,\nu_i,\tau_i),
\end{align*}
where $(\mu_i,\log\sigma_i,\nu_i,\log\tau_i)^\top
    =
    (\mathbf{x}_{\mu i}^{\top}\boldsymbol{\beta}_{\mu},
    \mathbf{x}_{\sigma i}^{\top}\boldsymbol{\beta}_{\sigma},
    \beta_{\nu0},
    \mathbf{x}_{\tau i}^{\top}\boldsymbol{\beta}_{\tau})^\top$. The model therefore uses identity links for $\mu$ and $\nu$, and log links for $\sigma$ and $\tau$. The shape parameter $\nu$ is specified as intercept-only, while covariates for the remaining distributional parameters are given in Table~\ref{tab:jsuo-m14-adj-final-svy}. The household wealth quintile variable (\texttt{V190}) is included as a categorical predictor with the middle quintile as reference category.

\begin{table*}[t]
\caption{Final JSUo GAMLSS model for child height-for-age z-scores with naive and survey-robust standard errors. The Change column compares significance at the $\alpha = 0.05$ level before and after replacing naive standard errors with survey-robust standard errors.}
\label{tab:jsuo-m14-adj-final-svy}
\centering
\small
\setlength{\tabcolsep}{3pt}
\renewcommand{\arraystretch}{1.06}

\begin{tabular*}{\textwidth}{@{\extracolsep{\fill}}clrccGrrcc@{}}
\toprule
Parameter & Term & Estimate & Naive SE & Survey-robust SE & \multicolumn{1}{c}{SE Ratio} & Model $p$ & Survey $p$ & Change & Sig.\\
\midrule
$\mu$ & Intercept & -1.9077 & 0.1948 & 0.2230 & 1.14 & $<$ 0.001 & $<$ 0.001 &  & ***\\
$\mu$ & V190Poorest & -0.3445 & 0.0554 & 0.0698 & 1.26 & $<$ 0.001 & $<$ 0.001 &  & ***\\
$\mu$ & V190Poorer & -0.2233 & 0.0555 & 0.0629 & 1.13 & $<$ 0.001 & $<$ 0.001 &  & ***\\
$\mu$ & V190Richer & 0.0830 & 0.0582 & 0.0712 & 1.22 & 0.155 & 0.244 &  & \\
$\mu$ & V190Richest & 0.5698 & 0.0679 & 0.0915 & 1.35 & $<$ 0.001 & $<$ 0.001 &  & ***\\
$\mu$ & V133\_y & 0.0306 & 0.0058 & 0.0070 & 1.20 & $<$ 0.001 & $<$ 0.001 &  & ***\\
$\mu$ & $\text{M14}^2$ & 0.0231 & 0.0078 & 0.0077 & 0.99 & 0.003 & 0.003 &  & **\\
$\mu$ & B4 & 0.2084 & 0.0355 & 0.0349 & 0.98 & $<$ 0.001 & $<$ 0.001 &  & ***\\
$\mu$ & B0 & -0.5360 & 0.1217 & 0.1689 & 1.39 & $<$ 0.001 & 0.002 &  & **\\
$\mu$ & HV009 & -0.0253 & 0.0120 & 0.0140 & 1.16 & 0.036 & 0.071 & No longer sig. & .\\
$\mu$ & child\_deaths & -0.1543 & 0.0586 & 0.0705 & 1.20 & 0.009 & 0.029 &  & *\\
$\mu$ & V102 & 0.1277 & 0.0566 & 0.0689 & 1.22 & 0.025 & 0.065 & No longer sig. & .\\
$\mu$ & age\_months & -0.0034 & 0.0014 & 0.0016 & 1.14 & 0.016 & 0.034 &  & *\\
$\mu$ & V012\_y & 0.0059 & 0.0033 & 0.0037 & 1.11 & 0.077 & 0.111 &  & \\
$\mu$ & HV014 & -0.0674 & 0.0345 & 0.0369 & 1.07 & 0.051 & 0.068 &  & .\\
$\mu$ & M14 & -0.0709 & 0.0374 & 0.0365 & 0.98 & 0.058 & 0.053 &  & .\\
\addlinespace[0.5em]
$\sigma$ & Intercept & 0.9444 & 0.1506 & 0.1615 & 1.07 & $<$ 0.001 & $<$ 0.001 &  & ***\\
$\sigma$ & child\_deaths & 0.0871 & 0.0397 & 0.0507 & 1.28 & 0.029 & 0.087 & No longer sig. & .\\
$\sigma$ & age\_months & -0.0022 & 0.0009 & 0.0009 & 1.05 & 0.017 & 0.022 &  & *\\
$\sigma$ & HV014 & 0.0458 & 0.0188 & 0.0240 & 1.28 & 0.015 & 0.057 & No longer sig. & .\\
$\sigma$ & c\_anem & 0.5110 & 0.5879 & 0.4544 & 0.77 & 0.385 & 0.261 &  & \\
$\sigma$ & V102 & 0.0580 & 0.0361 & 0.0530 & 1.47 & 0.109 & 0.275 &  & \\
$\sigma$ & V012\_y & 0.0033 & 0.0022 & 0.0025 & 1.15 & 0.137 & 0.195 &  & \\
$\sigma$ & m\_anem & 0.0274 & 0.0732 & 0.0749 & 1.02 & 0.709 & 0.715 &  & \\
\addlinespace[0.5em]
$\nu$ & Intercept & -0.1518 & 0.1508 & 0.1586 & 1.05 & 0.315 & 0.339 &  & \\
\addlinespace[0.5em]
$\tau$ & Intercept & 1.0624 & 0.1197 & 0.1328 & 1.11 & $<$ 0.001 & $<$ 0.001 &  & ***\\
$\tau$ & c\_anem & 0.4152 & 0.5605 & 0.4147 & 0.74 & 0.459 & 0.317 &  & \\
\botrule
\end{tabular*}
\end{table*}

\citet{meurerDeterminantsChildhoodMalnutrition2024} noted that GAMLSS did not provide a survey-robust variance estimator and that the resulting standard errors were therefore likely underestimated. We applied the proposed estimator to their final JSUo model and compare the resulting standard errors with the original model-based estimates in Table~\ref{tab:jsuo-m14-adj-final-svy}.

The standard-error corrections were considerably smaller than those observed in the most challenging simulation scenarios. This is consistent with the design characteristics of the application: the sample contained 3401 children from 500 PSUs and 60 strata, with approximately 6.8 observations per PSU. Moreover, several covariates capture household- and cluster-level variation, potentially reducing residual clustering in the score contributions. In contrast, the simulation scenarios were deliberately designed as stress tests with stronger cluster effects, larger within-cluster sample sizes, and more extreme weight variability. Hence, the moderate corrections observed in the Rwanda application are consistent with the simulation results.

Nevertheless, accounting for the survey design affected statistical inference. Four covariates that were significant under the naive standard errors no longer reached the 5\% significance level after correction. In particular, there was no longer sufficient evidence for associations between the outcome and household size (\texttt{HV009}) or urban/rural residence (\texttt{V102}). The latter finding should be interpreted in light of the strong association between residence and wealth status: richer households in Rwanda are more commonly located in urban areas, whereas poorer households are predominantly rural. After accounting for wealth quintiles, the remaining contribution of residence may therefore be insufficient to support a statistical association.

Similarly, the indicators \texttt{child\_deaths} (whether the mother experienced child loss) and \texttt{HV014} (number of siblings aged five or younger in the household) no longer showed sufficient evidence of an association with the variance parameter after using survey-robust standard errors.

\section{Discussion}

\subsection{Summary and conclusions}

Our results show that conventional model-based and model-robust GAMLSS standard errors perform well under simple random sampling but become severely anti-conservative once clustering and unequal weighting are introduced. The proposed survey-robust sandwich estimator provides well-calibrated standard errors across all simulation settings, including complex survey designs with clustering, stratification, and unequal probabilities of selection. The estimator performs close to nominal levels for all distributional parameters, with only minor deviations in the more challenging scenarios, particularly for the shape parameter $\nu$.

Replication-based methods, including bootstrap and delete-a-group Jackknife estimators, provide valid alternatives for survey inference in distributional regression models and generally achieved comparable calibration. However, their computational cost is substantially higher because they require repeated model fitting. This limitation becomes particularly relevant for flexible GAMLSS models, where likelihood optimisation and numerical derivatives can already be computationally demanding. In contrast, the proposed estimator requires only a single fitted model and therefore provides a computationally efficient approach for obtaining survey-robust standard errors.

\subsection{Limitations and future research}

The proposed estimator is based on a single-stage approximation of the survey variance. Although it performed well in the multi-stage sampling scenarios considered here, further improvements may be possible through extensions based on exact multi-stage variance estimators, such as those of \citet{horvitzGeneralizationSamplingReplacement1952} and \citet{yatesSelectionReplacementStrata1953}. In practice, however, the required higher-order inclusion probabilities are often unavailable due to confidentiality constraints, limiting the applicability of such approaches.

The slightly increased uncertainty in the estimation of the $\nu$ parameter warrants further investigation. It remains unclear whether this behaviour is driven by the characteristics of the synthetic populations considered here or by specific properties of the variance estimator. Additional simulation studies using alternative data-generating mechanisms and distributional families could provide further insight.

\noindent Finally, the current derivation focuses on parametric GAMLSS models. The broader GAMLSS framework allows more flexible predictors, including smooth terms, random effects, and other non-parametric components (\citealp[p.~112]{stasinopoulosGeneralizedAdditiveModels2024a}). Extending survey-robust variance estimation to these settings represents an important direction for future research. This challenge is not unique to GAMLSS; even for simpler models such as generalised additive models (GAMs; \citealp{hastieGeneralizedAdditiveModels1990}), fully developed design-based variance estimation remains an active area of research.

\section{Conflicts of interest}
The authors declare that they have no competing interests.

\section{Data availability}

The code and synthetic finite population required to reproduce the simulation study are archived on Zenodo\footnote{See \url{https://doi.org/10.5281/zenodo.20754402}.}. The latest development version of the implementation is available on GitHub\footnote{See \url{https://github.com/dmeurer94/survey-robust-gamlss}.}. The raw DHS microdata used in the real-data application are not publicly redistributed by the authors and must be obtained directly from the DHS Program in accordance with its data access and use policies.

\section{Author contributions statement}

D.M.\ conceived the idea, developed and implemented the methodology, conducted the simulation experiments, performed the data analysis, and interpreted the results. T.A.\ provided guidance throughout the project and critically revised the manuscript. D.M.\ and T.A.\ wrote and approved the final version of the manuscript.

\section{Acknowledgments}

The authors thank the DHS Program for providing access to the data used to construct the synthetic finite population underlying the simulation study and evaluation of the proposed estimator.

\section*{Use of AI tools}

During preparation of this manuscript, the authors used OpenAI's ChatGPT to support language editing, LaTeX and code troubleshooting, and manuscript organisation. The authors reviewed and edited all AI-assisted output and take full responsibility for the content of the manuscript, the code, and the reported results.

\bibliographystyle{oup-abbrvnat}
\bibliography{paper-refs}

\begin{appendices}

\section{Appendix}

\subsection{Regularity conditions}

The proposed estimator applies the implicit-estimator linearisation of
\citet{binderVariancesAsymptoticallyNormal1983} to the stacked GAMLSS
estimating equation. We therefore do not reproduce the general asymptotic
argument here. Instead, we state the conditions needed to place the present
GAMLSS estimator within that framework.


\noindent Let \(\mathcal{U}_N\) denote a sequence of finite populations and let
\(\boldsymbol{\theta}_N\) denote the finite-population root of 
\[
W_N^G(\boldsymbol{\theta}_N)=\mathbf{0}.
\]
The following conditions define the scope of the estimator:

\begin{enumerate}
  \item[(i)] \textbf{Finite-population target and fixed dimension.}
  The target \(\boldsymbol{\theta}_N\) is the finite-population root of the
  stacked GAMLSS estimating equation. The number of fitted coefficients
  \(p=\sum_m p_m\) is fixed along the asymptotic sequence.

  \item[(ii)] \textbf{Local identification and smoothness.}
  The target is an interior point of the parameter space, the finite-population
  root is locally unique, and the limiting derivative matrix is nonsingular.
  The individual log-likelihood contributions and inverse link functions are
  sufficiently smooth in a neighbourhood of \(\boldsymbol{\theta}_N\), and
  fitted values remain away from boundary points where derivatives fail to
  exist or the information matrix degenerates.

  \item[(iii)] \textbf{Design regularity for score totals.}
  The survey-weighted stacked score total satisfies the usual design
  consistency and central-limit conditions for implicit survey estimators.
  In particular, weighted unit-level score contributions and PSU-level score
  totals have bounded second moments, and no small number of units or PSUs
  dominates the design variance.

  \item[(iv)] \textbf{Consistent design-variance estimator.}
  Under the intended first-stage with-replacement cluster approximation, the
  stratified PSU-level variance estimator consistently estimates the design
  variance of the survey-weighted stacked score total.

  \item[(v)] \textbf{Fixed estimating equation and compatible weights.}
  The GAMLSS family, link functions, covariates and any smoothing or penalty
  choices are treated as fixed. The same analysis weights are used consistently
  in the estimating equation, the bread and the meat. A common multiplicative
  rescaling of all weights is harmless if applied throughout.
\end{enumerate}

\noindent Under these conditions, the regularity requirements of the
\citet{binderVariancesAsymptoticallyNormal1983} implicit-estimator framework
are met for the stacked GAMLSS score. The first-order design covariance is
therefore obtained by premultiplying and postmultiplying the design variance of
the survey-estimated stacked score total by the inverse bread matrix and its
transpose. The proposed plug-in estimator follows by replacing the
finite-population bread and score-total variance by their sample analogues.

\subsection{Special cases and checks}
\label{appendix:special_cases}

The construction reduces to familiar survey-sandwich estimators in several special cases.

\subsubsection{Reduction to a one-parameter GLM}
If the fitted GAMLSS has only one distributional parameter block, then $M=1$, $\boldsymbol{\theta}=\boldsymbol{\beta}$, and the stacked score has a single component. Writing the GLM linear predictor as
\[
    \eta_i=x_i^{\top}\boldsymbol{\beta},
\]
the unit GAMLSS score reduces to
\[
    u_i^G(\boldsymbol{\beta})
    =
    \frac{\partial \ell_i(\boldsymbol{\beta})}{\partial \boldsymbol{\beta}}
    =
    x_i q_i(\boldsymbol{\beta}),
    \qquad
    q_i(\boldsymbol{\beta})
    =
    \frac{\partial \ell_i(\boldsymbol{\beta})}{\partial \eta_i}.
\]
For a standard GLM, $q_i(\boldsymbol{\beta})$ is proportional to the usual residual score, so the weighted estimating equation becomes
\[
    \widehat W_s^G(\boldsymbol{\beta})
    =
    \sum_{i\in s} a_i x_i q_i(\boldsymbol{\beta})
    =
    0.
\]
The corresponding bread and PSU-level score totals are
\[
    \widehat B_s^G
    =
    -
    \sum_{i\in s}
    a_i x_i x_i^{\top}
    \frac{\partial q_i(\boldsymbol{\beta})}{\partial \eta_i}
    \bigg|_{\boldsymbol{\beta}=\widehat{\boldsymbol{\beta}}},
    \qquad
    \widehat T_{hj}
    =
    \sum_{i\in s_{hj}}
    a_i x_i \widehat q_i .
\]
Substituting these quantities into
\[
    \widehat{\operatorname{Var}}_p(\widehat{\boldsymbol{\beta}})
    =
    \left(\widehat B_s^G\right)^{-1}
    \widehat \Omega_{\mathrm{WR}}^G
    \left(\widehat B_s^G\right)^{-\top}
\]
gives the usual Binder-type survey sandwich for GLM estimating equations \citep{binderVariancesAsymptoticallyNormal1983}. Thus, the proposed estimator is a direct extension of the GLM survey sandwich from a single residual-score equation to the stacked score equations of a multi-parameter GAMLSS.

\subsubsection{Unclustered sample}
If there is no clustering, each observation can be treated as its own PSU. The PSU-level score total is then
\[
\mathbf{T}_{hi}=a_i\mathbf{u}_i(\widehat{\boldsymbol{\theta}}),
\]
where $a_i$ is the analysis weight and $\mathbf{u}_i(\widehat{\boldsymbol{\theta}})$ is the individual GAMLSS score contribution evaluated at the fitted parameter vector. The meat then becomes the stratified design variance estimator for a weighted total of individual score contributions.

\subsubsection{No stratification}
If there is no stratification, take $H=1$. The with-replacement cluster meat reduces to
\[
\widehat{\mathbf{\Omega}}_{\mathrm{WR}}
=
\frac{m}{m-1}
\sum_{j=1}^{m}
\left(\mathbf{T}_{j}-\overline{\mathbf{T}}\right)
\left(\mathbf{T}_{j}-\overline{\mathbf{T}}\right)^{\top},
\]
where $m$ is the number of sampled PSUs, $\mathbf{T}_j$ is the weighted score total for PSU $j$, and $\overline{\mathbf{T}}=m^{-1}\sum_{j=1}^{m}\mathbf{T}_j$.

\subsubsection{Invariance to constant weight rescaling}
Suppose all weights are multiplied by a constant $c>0$, and the rescaled weights are used consistently in the fitted estimating equation, bread, and meat. Then
\[
\widehat{\mathbf{B}}\mapsto c\widehat{\mathbf{B}},
\qquad
\widehat{\mathbf{\Omega}}\mapsto c^2\widehat{\mathbf{\Omega}}.
\]
Therefore,
\[
(c\widehat{\mathbf{B}})^{-1}
(c^2\widehat{\mathbf{\Omega}})
(c\widehat{\mathbf{B}})^{-\top}
=
\widehat{\mathbf{B}}^{-1}
\widehat{\mathbf{\Omega}}
\widehat{\mathbf{B}}^{-\top}.
\]
The final sandwich covariance estimator is therefore invariant to a common multiplicative rescaling of all weights, provided the rescaling is used consistently throughout the estimator.

\begin{table}[h]
\small
\caption{Number of active non-intercept regressors by simulation scenario, GAMLSS family, and distributional parameter. A dash indicates that the parameter is not part of the fitted family in that scenario, whereas 0 indicates an intercept-only parameter.}
\label{tab:term_counts}
\centering
\begin{tabular*}{\columnwidth}{@{\extracolsep{\fill}}cccccc@{}}
\toprule
Scenario & Family & $\mu$ terms & $\sigma$ terms & $\nu$ terms & $\tau$ terms \\
\midrule
S1  & Normal & 6 & 0 & -- & -- \\
S2  & Normal & 6 & 3 & -- & -- \\
S3  & BCPEo  & 6 & 3 & 2  & 0  \\
S4  & BCPEo  & 8 & 4 & 3  & 2  \\
\addlinespace[0.3em]
S5  & BCPEo  & 8 & 4 & 3  & 2  \\
S6  & BCPEo  & 8 & 4 & 3  & 2  \\
S7  & BCPEo  & 8 & 4 & 3  & 2  \\
S8  & BCPEo  & 8 & 4 & 3  & 2  \\
S9  & BCPEo  & 8 & 4 & 3  & 2  \\
S10 & BCPEo  & 8 & 4 & 3  & 2  \\
\botrule
\end{tabular*}
\end{table}

\begin{figure*}[h]
\centering
\includegraphics[
  width=\textwidth,
  height=0.82\textheight,
  keepaspectratio
]{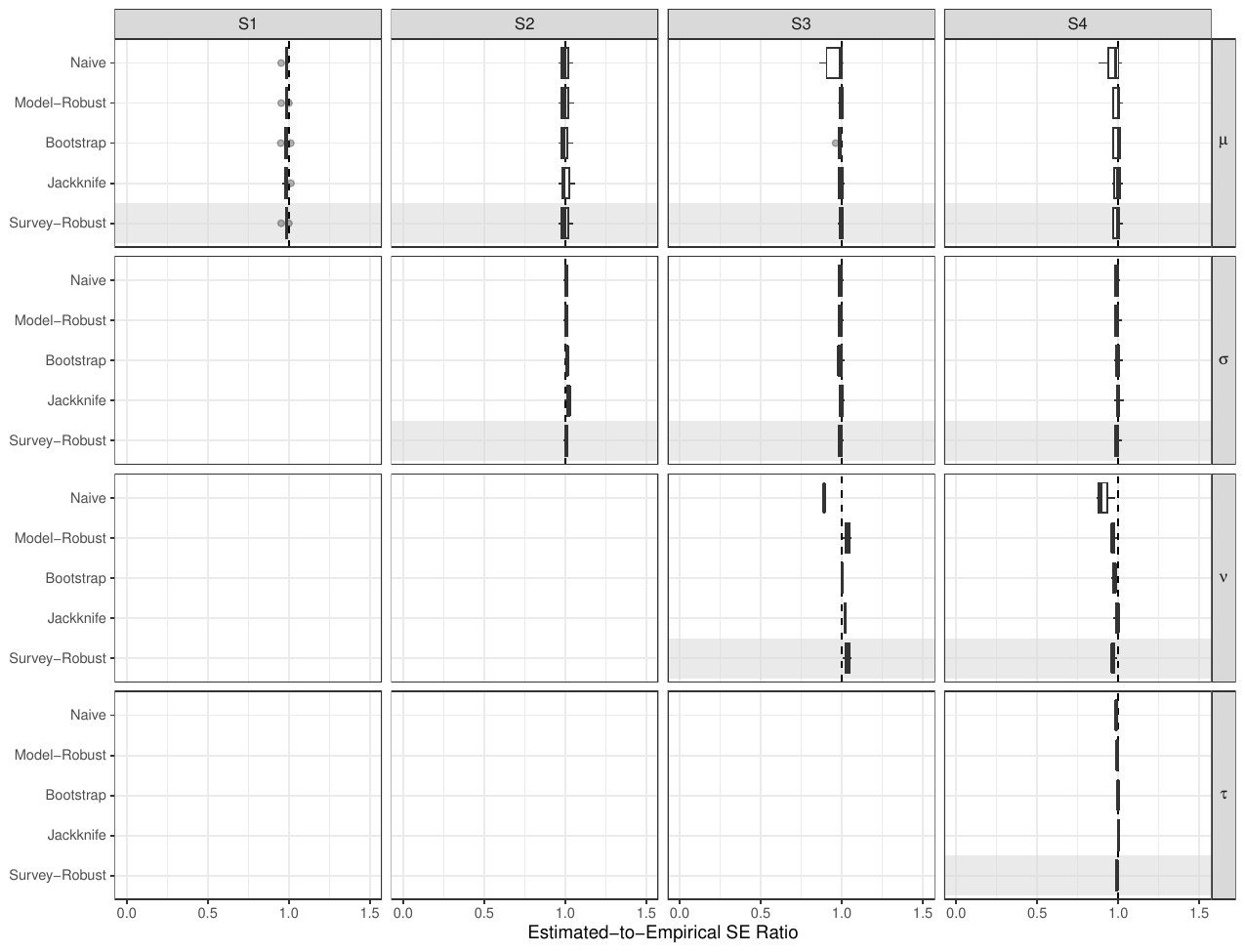}
\caption{Coefficient-level distributions of estimated-to-empirical standard-error ratios for scenarios S1--S4. Boxplots are shown separately by GAMLSS parameter, scenario, and variance estimator. Shaded in grey is the proposed survey-robust sandwich estimator. Intercepts are excluded. The dashed vertical reference line indicates perfect calibration. Values below one indicate underestimation of standard errors, while values above one indicate conservative standard errors.}
\label{fig:appendix_se_ratio_boxplots_1}
\end{figure*}

\end{appendices}

\end{document}